\documentclass[english,ngerman]{bvm} 

\bvmdef\articlenumber{0000}

\bvmdef\type{V}

\date{}

\title{Detection of Large Vessel Occlusions using Deep Learning by Deforming Vessel Tree Segmentations}

\titlerunning{LVO Detection using DL and Deformation-Based Augmentation}

\author{Florian Thamm \inst{1,2} \and Oliver Taubmann \inst{2}\and Markus J\"urgens \inst{2}\and Hendrik Ditt \inst{2}\and Andreas Maier \inst{1}}
\authorrunning{F. Thamm et al.}
\institute{Friedrich-Alexander University Erlangen-Nuremberg, Erlangen, Germany \and Siemens Healthcare GmbH, Forchheim, Germany}

\email{florian.thamm@fau.de}

\begin{document}

\selectlanguage{english}

\maketitle

\begin{abstract}
 Computed Tomography Angiography is a key modality providing insights into the cerebrovascular vessel tree that are crucial for the diagnosis and treatment of ischemic strokes, in particular in cases of large vessel occlusions (LVO). Thus, the clinical workflow greatly benefits from an automated detection of patients suffering from LVOs. This work uses convolutional neural networks for case-level classification trained with elastic deformation of the vessel tree segmentation masks to artificially augment training data. Using only masks as the input to our model uniquely allows us to apply such deformations much more aggressively than one could with conventional image volumes while retaining sample realism.
 The neural network classifies the presence of an LVO and the affected hemisphere. In a 5-fold cross validated ablation study, we demonstrate that the use of the suggested augmentation enables us to train robust models even from few data sets. Training the EfficientNetB1 architecture on 100 data sets, the proposed augmentation scheme was able to raise the ROC AUC to 0.85 from a baseline value of 0.56 using no augmentation. The best performance was achieved using a 3D-DenseNet yielding an AUC of 0.87. The augmentation had positive impact in classification of the affected hemisphere as well, where the 3D-DenseNet reached an AUC of 0.93 on both sides.
\end{abstract}

\section{Introduction}
Computed Tomography Angiography (CTA) is a commonly used modality in many clinical scenarios including the diagnosis of ischemic strokes. An ischemic stroke is caused by an occluded blood vessel resulting in a lack of oxygen in the affected brain parenchyma. An occlusion in the internal carotid artery (ICA), proximal middle cerebral artery (MCA) or basilar artery is often referred to as large vessel occlusion (LVO). These LVOs are visible in CTA scans as a discontinuation of contrast agent in the vascular tree, which is a complex system of arteries and veins and varies from patient to patient. Consequently the diagnosis takes time and requires expertise. Clinics and patients would therefore benefit from an automated classification of LVOs on CTA scans. 

Prior research in that field is described in literature. Amukotuwa et al.~detected LVOs in CTA scans with a pipeline consisting of 14 steps and tested their commercially available algorithm on two different data cohorts, reporting a performance of 0.86 ROC AUC in the first trial with 477 patients \cite{amukotuwa2019automated} and 0.94 in the second trial \cite{amukotuwa2019fast} with 926 patients. Stib et al.~\cite{stib2020detecting} computed maximum-intensity projections of segmentations of the vessel tree based on multi-phase CTA scans (three CTA scans covering the arterial, peak venous and late venous phases), and trained a 2D-DenseNet \cite{huang2017densely} on 424 patients to classify the presence of an LVO. They report ROC AUC values between 0.74 and 0.85 depending on the phase. Luijten et al.'s work \cite{luijten2021diagnostic} investigated the performance of another commercially available LVO detection algorithm based on a Convolutional Neural Network (CNN) and determined a ROC AUC of 0.75 on 646 test patients.

In all studies, very large data cohorts were available. This appears mandatory to train (and test) AI-based detection algorithms, since in case-wise classification the number of training samples equals the number of available patients. In this work we present a data-efficient method that achieves performance comparable to what is seen in related work while relying on only 100 data sets for training.

\section{Materials and methods}
\subsection{Data}
Altogether, 168 thin-sliced ($0.5$ to $1$\,mm) head CTA data sets were available. Of these, 109 patients were LVO positive due to an occlusion either in the middle cerebral artery or the internal carotid. Regarding the affected hemisphere, 54 (52) LVOs were located on the left (right) side. The data was acquired from a single site with a Somatom Definition AS+ (Siemens Healthineers, Forchheim, Germany).
\subsection{Methodology}
\begin{figure}[b]
	\centering
	\caption{Proposed pipeline. First, the CTA volume is registered to a probabilistic brain atlas. The cerebrovascular vessel tree is then segmented and transformed into the atlas coordinate system by applying the resulting deformation vector field. For augmentation purposes, the vessel tree segmentation masks are elastically deformed for training only. A network predicts the three mutually exclusive classes: No LVO, left or right LVO.}
     \label{fig:pipe}
    \includegraphics[width = \textwidth]{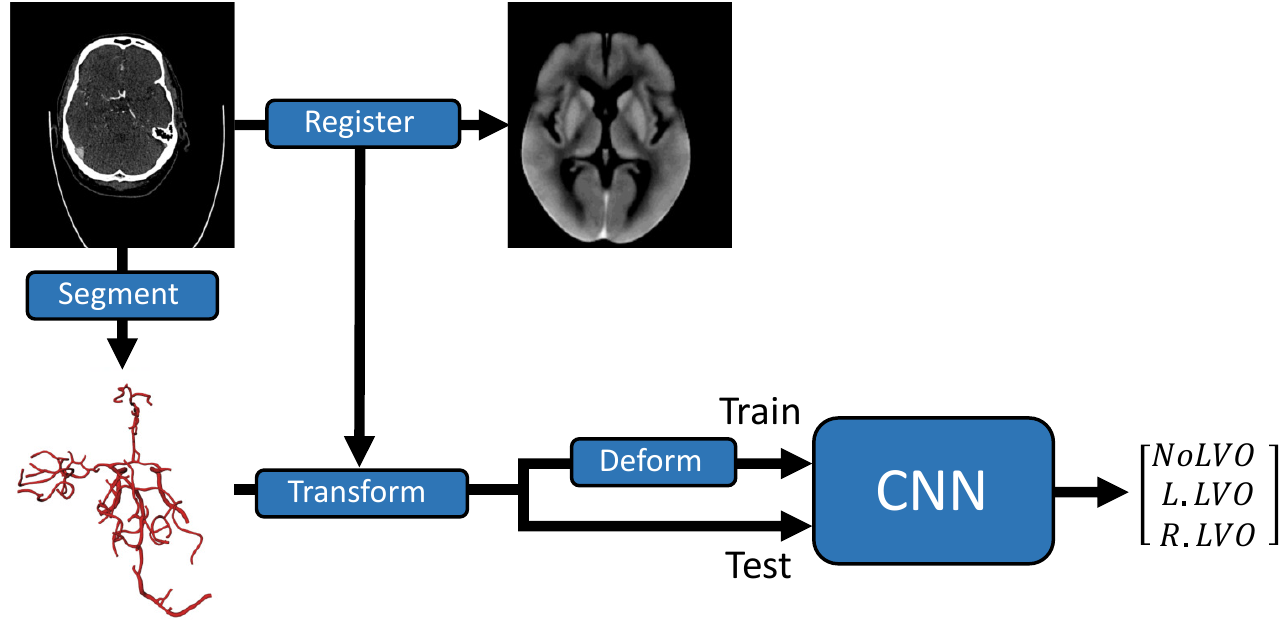}
	
\end{figure}

The method we propose is based on the idea of aggressively augmenting the vessel tree segmentations in order to artificially extend the amount of trainable data. The classification pipeline itself (Fig.~\ref{fig:pipe}), consists of three subsequent steps. In the first step, the cerebrovascular tree is segmented using segmentation approach published by Thamm et al.~\cite{thamm2020virtualdsa}. Additionally, the algorithm prunes the vessel tree to the relevant arteries by masking out all vascular structures which are a walking distance (geodesic distance w.r.t.~vessel center lines) of more than 150\,mm away from the Circle of Willis. Thereby veins, which are not relevant in the diagnosis of LVOs, like the Sinus Sagittalis are mostly excluded from further processing. In the second step, the original CTA scan is non-rigidly registered to the probabilistic brain atlas by Kemmling et al. \cite{kemmling2012decomposing}. The registration is based on Chefd'hotel et al.'s method \cite{chefd2002flows} but may be done using other, publicly available registration methods as well. The resulting deformation field is used to transform the segmentation mask into the atlas coordinate system. An accurate registration between an atlas and the head scan is not crucial in our work as  variations in the vessel tree are present in all patients anyway. Once the volumes are in the atlas coordinate space, they are equally sized with 182 $\times$ 205 $\times$ 205 voxels with isotropic spacing of 1\,mm in all dimensions. The primary purpose of the registration is to consistently orient and somewhat anatomically ``normalize'' the segmentation for the next step, in which a convolutional neural network classifies the presence of an LVO. The network receives the binary segmentation masks volume-wise and predicts a softmax activated vector of length 3, representing the three classes: No LVO, left LVO and right LVO. In our work, we tested a 3D-version of DenseNet \cite{hara3dcnns} ($\approx$ 4.6m parameters) and EfficientNetB1 \cite{tan2019efficientnet} ($\approx$ 6.5m parameters) where the channel dimension has been repurposed as the z-Axis. Cross entropy serves as the loss optimized with Adam on PyTorch 1.6 \cite{pytorch} and Python 3.8.5.
 
\subsection{Augmentation}
\label{sec:augmentation}
 From patient to patient, the cerebrovascular anatomy follows coarsely the same structure. However, anatomical variations (e.g.,~absence ICA, accessory MCA) combined with the individual course of the vessels lead to a wide variety of configurations in intracranial vascular systems such that no vessel tree is quite like another. Considering this, augmentations can be used in order to artificially generate more vessel trees and from a network's perspective visually new patients. Therefore, we propose to elastically and randomly deform the segmentation masks for training.
 While the use of elastic deformation for augmentation per se is not a novel technique \cite{nalepa2019data}, in our setup we are uniquely able to apply it much more aggressively than otherwise possible, enabling us to dramatically increase its benefit compared to typical use cases.
 This is possible due to the fact that only vessel segmentations are used as input for our CNN-based classifier model. Whereas strong deformations on a conventional image volume will quickly introduce resampling artifacts that render the image unrealistic, masks remain visually comparable to the original samples even when heavily deformed. As the segmentation is performed on full volumes, an online augmentation, i.e.~deforming while the network is trained, is computationally too expensive and would increase the training time to an impractical level. Instead, we suggest to elastically deform the segmentation masks prior to the training for a fixed number of random fields for each original volume.
 As masks, unlike regular image volumes, are highly compressible, this does not notably increase data storage requirements as would typically be associated with such an approach.
 
\begin{figure}[b]
	\centering
	    \caption{Two examples with four augmentations of the original tree on the left, all viewed axially caudal with the same camera parameters. The upper row shows a case with an occlusion in the left middle cerebral artery, indicated by the arrow. The lower row shows an LVO-negative vessel tree. Instead of binary masks, surface meshes of the deformed masks were rendered for a sharper and clearer visualization.}
     \label{fig:deform}
    \includegraphics[width = \textwidth]{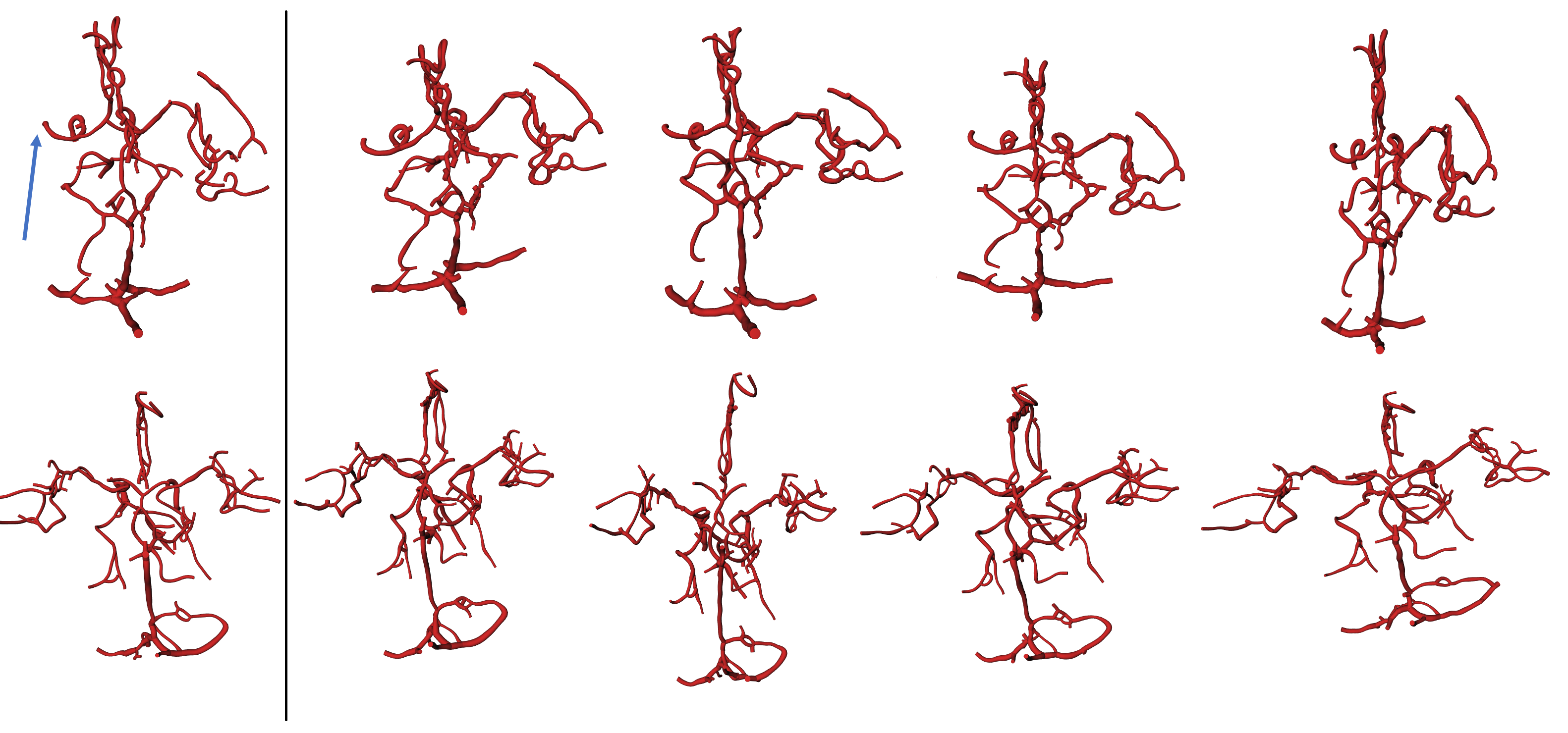}

\end{figure}

 In this work we aim to demonstrate the impact of this data augmentation on the performance for the classification of LVOs. Using the RandomElasticDeformation of TorchIO \cite{torchio} which interpolates displacements with cubic B-splines, we randomly deformed each segmentation mask 10 times with 4 and 10 times with 5 random anchors, all 20 augmentations with a maximal displacement of 90 voxels (Examples in Fig.~\ref{fig:deform}). Additionally, we mirror the original data sets sagitally and again apply the above procedure to create 20 variants, which flips the right/left labels but has no effect if no LVO is present. We thus create 40 samples out of one volume, resulting in a total of 6720 vessel tree samples generated from 168 patients. 
 
\section{Results}
We investigated the impact on the elastic augmentations in an ablation study considering two architectures, where we systematically disable features (deformation and mirroring). For a fully 3D variant we evaluated the 3D-DenseNet architecture \cite{hara3dcnns} and a 2D variant where the channel axis of the input is used for the axial ($z$) dimension using the EfficientNetB1 architecture \cite{tan2019efficientnet}. A 5-fold cross validation setup with a 3-1-1 split ratio for training, validation and testing was conducted, where on average, 100 original data sets were used for training per cycle. The baseline (no augmentation) was trained for 200 epochs, a variant using the original and the deformed, but not mirrored data sets was trained for 100 epochs, and finally, as the proposed setting, models were trained for 50 epochs using the original, the deformed and mirrored data. Epoch numbers differ as there are more samples available when augmentation is used. All models overfitted by the end of their allotted epochs. The validation loss was used to pick the best performing network out of all epochs. The test data was not augmented to provide a fair comparison between all setups. Both architectures significantly benefit from the deformation-based augmentation (Tab.~\ref{tab:results}); in particular, EfficientNet failed to grasp the problem at all without it. The 3D-DenseNet trained with deformed and mirrored data sets outperformed the other setups by a significant margin, especially in detecting LVOs and left LVOs. Depending on the chosen threshold, this variant achieved a sensitivity of 80\% (or 90\%) and a specificity of 82\% (or 60\% respectively) for the detection of LVOs.   

\begin{table}[t]

\caption{ROC AUCs with 95\% confidence intervals (by bootstrapping) for the 3D-DenseNet and EfficientNetB1 architecture. ``D'' stands for ``deformation'' and ``M'' for ``mirroring''. To compute ``AUC Left'', the right and no-LVO class were combined to one class enabling a binary classification. ``AUC Right'' was calculated analogously.}
\label{tab:results}
\begin{tabular*}{\textwidth}{l@{\extracolsep\fill}lll}
\hline
Setup                  & AUC LVO             & AUC Left   & AUC Right  \\ \hline
3D-DenseNet + D + M     & \textbf{0.87} {[}0.81, 0.92{]} & \textbf{0.93} {[}0.87, 0.97{]} & 0.93 {[}0.88, 0.97{]} \\
3D-DenseNet + D         & 0.84 {[}0.77, 0.90{]} & 0.89 {[}0.84, 0.94{]} & \textbf{0.94} {[}0.90, 0.97{]} \\
3D-DenseNet             & 0.77 {[}0.69, 0.84{]} & 0.85 {[}0.78, 0.91{]} & 0.85 {[}0.78, 0.92{]} \\
EfficientNetB1 + D + M & 0.85 {[}0.79, 0.90{]} & 0.86 {[}0.79, 0.91{]} & 0.90 {[}0.84, 0.96{]} \\
EfficientNetB1 + D     & 0.83 {[}0.77, 0.89{]} & 0.85 {[}0.79, 0.91{]} & 0.91 {[}0.85, 0.96{]} \\
EfficientNetB1         & 0.56 {[}0.47, 0.65{]} & 0.59 {[}0.49, 0.68{]} & 0.68 {[}0.58, 0.78{]} \\ \hline
\end{tabular*} 
\end{table}
\section{Discussion}
We presented a method for automated classification of LVOs based on CTA data which makes heavy use of deformation fields for augmentation. With an AUC of 0.87 for LVO detection, we achieved a performance comparable to that of other DL-based approaches while using as few as 100 patient data sets for training. While not novel in itself, elastic deformation for the purpose of augmentation could be applied much more aggressively in our setup compared to regular use cases as our model relies exclusively on segmented vessel tree masks as input; for these, even strong deformations---that would cause severe resampling artifacts when applied to regular image volumes---still lead to anatomically meaningful representations that are virtually indistinguishable from real samples. In an ablation study we showed that the performed augmentation was crucial to properly learn the task at hand from a small number of data sets. This leads us to the conclusion that a learning-based detection of LVOs stands and falls with the number of training data sets. The cerebrovascular system is highly patient-specific, which is why the use of sophisticated augmentation techniques offers great potential. We postulate that also larger data pools could benefit from more extensive data augmentation if applied meaningfully.

\bibliographystyle{bvm}

\bibliography{3214}

\marginpar{\color{white}E\articlenumber}

\end{document}